\documentclass[a4paper]{article}

\usepackage{multirow}
\usepackage{authblk}

\usepackage{graphicx}
\usepackage[squaren]{SIunits}
\usepackage{color}

\usepackage[utf8]{inputenc}
\usepackage[T1]{fontenc}

\newcommand{\gfrac}[2]{\displaystyle\frac{#1}{#2}}

\newcommand{\dd}{\mbox{d}}
\newcommand{\eff}{\mathrm{eff}}
 
\begin{document}


\title{$\gamma$-ray polarimetry with conversions to $e^+e^-$
 pairs: polarization asymmetry and the way to measure it}

\author[1]{P.~Gros}
\author[1]{D.~Bernard\thanks{denis.bernard at in2p3.fr}}
\affil[1]{LLR, \'Ecole Polytechnique, CNRS/IN2P3, 91128 Palaiseau, France}

\maketitle

\begin{abstract}
We revisit the measurement of the polarization fraction, $P$, and the
measurement of the polarization angle of partially linearly-polarized gamma
rays using their conversion to $e^+e^-$ pairs in the field of a
nucleus.
We show that an inappropriate definition of the azimuthal angle,
$\varphi$, used to reference the orientation of the final state
degrades the precision of the measurement of $P$, by
comparison to the optimal case where the bisector angle of the electron and of
the positron momenta is used.
We then focus on the lowest part of the energy spectrum, below
$\approx 10\,\mega\electronvolt$, where a large part of the statistics
lie for a cosmic source.
We obtain the value of the polarization asymmetry, $A$, of pair conversion
at threshold and we show that in the case where the correct
expression is used for $\varphi$, the measured value of $A$ tends
to the limit.
\end{abstract}




\section{Introduction}
\label{sec:introduction}

\subsection{$\gamma$-ray polarimetry and astrophysics}
\label{subsec:polar:et:astro}

While at low energies (from radio waves to X-rays),
polarimetry is a powerful tool to gain an insight into understanding
the working mechanisms of cosmic sources, this diagnostic is missing for photon
energies $E > 1\,\mega\electronvolt$.
Polarimetry would provide information on the magnitude, on the direction and on
the structure, in particular on the turbulence, of the magnetic fields
in the regions that are emitting high-energy $\gamma$ rays, in particular
of relativistic jets.

{\bf Blazars}, for example, active galactic nuclei in which one jet 
 points at a small angle with respect to our line of sight,
show SEDs (spectral energy distributions) that have two broad maxima,
both originating from non-thermal emission:
The low-energy component is believed to proceed from synchrotron
radiation from relativistic electrons, while the origin of the
high-energy (X-ray through $\gamma$-ray) component is still under
debate, as both leptonic and hadronic models produce acceptable fits
to the SEDs of most blazars (a recent review can be found in this
discussion of the 2010 flare of 3C 454.3 \cite{Diltz:2016aqg}).
In the leptonic model, the high energy component has
contributions from synchrotron-self-Compton (SSC) and from external Compton
(EC) radiation whose seed photons are from direct accretion disk
emission and from any other external radiation field and are un-polarized.
In the hadronic model, the high-energy emission consists primarily of
contributions from proton synchrotron emission and synchrotron
emission from secondary pairs produced in cascade processes.
In most blazar classes, 
hadronic and leptonic models show similar degrees of X-ray 
polarization which makes polarization a weakly discriminant
diagnostic, while in the $\gamma$ energy range hadronic models predict
still a large polarization fraction (up to $P \approx 70\%$), in contrast
with leptonic models for which the small polarized SSC ``signal'' is
 washed out by the huge unpolarized EC emission component: high-energy
$\gamma$-ray polarimetry will yield the discriminating diagnostic
\cite{Zhang:2013bna,Chakraborty:2014vma}.
A demonstration that a hadronic mechanism is at work in blazars would
make AGNs the preferred suspect for the long-sought acceleration site
of ultra-high-energy cosmic rays (UHERC).

For {\bf pulsars}, after Fermi-LAT's study showed that high-altitude
emission zones are favored by observations\,\cite{Abdo:2009ax}
thereby disfavoring polar-cap models\,\cite{Dyks:2004ci}, one cannot
cleanly discriminate between
the two-pole caustic ``slot gap''\,\cite{Dyks:2004ci,Cerutti:2016hah},
the outer gap\,\cite{Dyks:2004ci,Takata:2007ut,Takata:2006cv} and
the striped wind\,\cite{Petri:2005ys,Petri:2013esa}
models, in particular as the pulsar inclination and viewing angles are
nuisance parameters.

In contrast, polarization can provide a handle on the emission
model, as the polarization signatures differs for the different
models\,\cite{Dyks:2004ci,Petri:2005ys,Takata:2007ut,Takata:2006cv,Petri:2013esa,Cerutti:2016hah,Harding:NASA2016,Harding:Kalapotharakos}.
As the polarization of the synchrotron emission is perpendicular to
that of curvature radiation and as the transition between synchrotron
and curvature radiation is expected to take place in the range 
1--100\,MeV 
in Crab-like pulsars\,\cite{Harding:NASA2016,Harding:Kalapotharakos}, a phase swing
is expected to be observed by MeV polarimeters, a signature for that
transition.
Theoretical prediction of the energy-dependent polarization fraction
and polarization angle phasograms from optical to high-energy
$\gamma$-rays will be needed to use these results for model
discrimination, see, e.g., Fig.\,6 of Ref.\,\cite{Takata:2007ut}.

Note also the recent quantum electrodynamics (QED) treatment of
curvature radiation in pulsar magnetospheres after which spin-flip
radiation makes an important contribution in (Crab-like) young pulsars
and it even becomes dominant (with respect to constant-spin
(classical) emission) in magnetars, and
contrary to constant-spin (classical) radiation, radiation
is mostly unpolarized\,\cite{Voisin:gamma2016,Voisin:2016}.

\subsection{$\gamma$-ray polarimetry and fundamental science}
\label{subsec:polar:et:science:fondamentale}

We have many reasons to think that the Standard Model (SM) could be
the low-energy remnant of a broken high-energy symmetry.
To parametrize this unknown, the Standard-Model Extension (SME) has
been built\,\cite{dcak,akgravity}, an effective field theory (EFT)
that contains the Standard Model, general relativity, and, ordered
after their mass dimension $d$, all possible operators that violate
Lorentz symmetry ({\bf Lorentz Invariance Violation}, LIV)
\footnote{Since $d=1$ operators are absent in a linear theory and
 $d=2$ operators are gauge-violating, the development includes
 operators with $d \ge 3$. $d$-even operators conserve the CPT
 symmetry, while $d$-odd operators violate it\,\cite{Kostelecky:2008ts}.}, be it a global
\cite{dcak} or a local \cite{akgravity} LIV.
In the photon sector, SME operators produce a number of effects that
can be tested experimentally\,\cite{Kostelecky:2008ts}.
Dispersion is the variation of the speed of light with photon energy and is due to $d \ne 4$ operators.
Birefringence arises from the fact that the operator parameters for
left and right circular polarized photons can have opposite signs: any
linear polarization is therefore rotated through an energy-dependent
angle as photons propagate, which depolarizes an initially linearly
polarized radiation consisting of a range of photon energies.
Hence, from the observation of linearly polarized radiation coming
from a distant source we obtain upper limits of the operator parameter
values.
Most birefringence tests are much more sensitive than dispersion
tests, so only when no birefringence is present does dispersion
provide useful information as is the case for $d=6$ limits, obtained
from the non observation of any time-of-flight/energy correlations in
AGN flares
(of Mrk 501 by MAGIC\,\cite{Albert:2007qk},
 of PKS 2155-304 by HESS\,\cite{Aharonian:2008kz}) nor in
GRB090510 by the Fermi-LAT\,\cite{Vasileiou:2013vra}.
For $d = 3$, limits are based on cosmic microwave background (CMB)
polarization studies, while for $d=4$ and $d=5$, birefringence limits
are obtained from measurements of the polarization fraction
of the 
soft-$\gamma$-ray emission of GRBs (in the 200--325 keV and in the
70-300 keV energy ranges for integral/IBIS and IKAROS, respectively)
\cite{Kostelecky:2008ts}.
As none of these soft-$\gamma$-ray GRB polarimetry measurements, taken
alone, is found to be statistically significant (see the review in
Ref.\,\cite{Stecker:2011ps}) and as the rotation angle is proportional
to the square of the photon energy, $E^2$, the development of
higher-energy polarimeters is eargerly needed.

Polarimetry of the radiation emitted by a far away source enables the
search for hints of the {\bf axion}\,\cite{Rubbia:2007hf}, the
putative
pseudoscalar pseudo-Goldstone boson induced by the breaking of the
$U(1)$ symmetry devised to solve the QCD CP problem.
Due to the axion-photon coupling, the propagation of the photon
through the magnetic field generated by the GRB would induce a
birefringence, that is a rotation of the direction of polarization,
that turns out to depend on the photon energy, so that the effective
average polarization fraction after propagation of a photon beam having a
sizable energy spread would be diluted.
The observation of a non-zero polarization fraction then translates to 
 an upper limit on the axion-photon coupling
 $g_{a\gamma\gamma}$.
 The limit is GRB-model dependent, but it is presently the best limit for an
axion mass close to $1\,\milli\electronvolt$\,\cite{Rubbia:2007hf} 
(compare to the present Fig.\,1 of Ref.\,\cite{pdg:2016}).
As the limit is proportional to $1/\sqrt{E}$, extending the
polarization measurement to higher energies would lead to an improved
value, or even to a detection of a possible axion-like-particle (ALP),
one of these pseudoscalars that are not bound to the mass-to-coupling-constant
relation that axions are subject to, and that may or may not be a
component of dark matter.

\subsection{$\gamma$-ray polarimetry with pairs: techniques}
\label{subsec:polar:techniques}

In contrast with the radio-wave and optical regimes for which
polarimetry is performed by the measurement of electric fields or
of light intensities, in the X-to-$\gamma$-ray regime photons are
detected one by one:
the polarization information is extracted from a sample of such
conversion events, from the distribution of an angle that is a measure
of the orientation of the final state particles with respect to the
polarization direction of the photon, an azimuthal angle, and that is
denoted generically $\varphi$ in this paper.

Whatever the process at work during conversion, be it photo-electric effect,
Compton scattering or pair conversion, due to the $J^{PC} = 1^{--}$
quantum numbers of the photon, the differential cross section is
described by the expression:
\begin{equation}
\gfrac{\dd \sigma}{\dd \varphi} \propto 
\left(
1 + A \times P \cos(2(\varphi - \varphi_0))
\right).
\label{eq:modulation}
\end{equation}

The modulation factor of the cosine, $A \times P$, is the product of
the polarization asymmetry of the conversion process (here of pair
conversion), $A$, and of the linear polarization fraction of the
incoming radiation, $P$.
Both $A$ and $P$ are defined to be in the range $[0,1]$ and $\varphi_0$
is the polarization angle of the incoming radiation.
Experimental effects affect the measurement of $\varphi$, which leads to
an effective asymmetry, $A_{\eff}$ that is lower than the QED
asymmetry, $A$.
Their ratio $D \equiv A_{\eff}/A$ is named the asymmetry dilution due
to the experimental effects, $0 \le D \le 1$.

Measuring $\gamma$-ray linear polarization by pair production was
first suggested by Yang\,\cite{Yang1950} and
the full polarized differential cross section was given by 
Berlin\,\cite{BerlinMadansky1950} and May\,\cite{May1951}
(an exhaustive review can be found at Ref.\,\cite{Motz:1969ti}).
Unfortunately, the {\bf multiple scattering} of the two lepton tracks in the high-$Z$ material converter plates of the past (COS-B, EGRET) and present (Fermi-LAT) telescopes blurs the azimuthal information, 
to the extent that the measurement of the azimuthal angle is impossible
\cite{Kelner,Kotov,Mattox,Yadigaroglu:1996qq},
the key point being that the lower multiple scattering undergone by
higher-momentum tracks from the conversion of higher-energy photons is
compensated for by the fact that higher-energy photons convert to pairs
with a smaller average opening angle\footnote{In Si/W trackers, in the case the wafers are located
 downstream to the W converter, there seems to be some hope that the
 selection of conversion events that took place in the silicon enable the 
 measurement of the azimuthal angle with some
 precision\,\cite{Giomi:2016brf}.}.

A way out has been sought with {\bf triplet conversions}, in which the
target electron recoils at a large polar angle, giving hope that
measuring its azimuthal angle would be easier
\cite{Boldyshev:1994bs,Depaola:2009zz,Iparraguirre:2011zz}.
Unfortunately, the useable fraction of the cross section, that is, the cross
section of triplet conversion with the electron recoil momentum large
enough that the track can be reconstructed is extremely small 
(Fig.\,6 of Ref.\,\cite{Bernard:2013jea}), even for gas detectors,
with the consequence that the sensitivity of a space polarimeter that
would use triplet conversion is very low 
(section 5.3 and Figs. 25-26 of Ref.\,\cite{Bernard:2013jea}).

Past theoretical works showed that the asymmetry is larger when the
two leptons share the energy equally\,\cite{Olsen:1959zz},
that it varies with the (azimuthal) acoplanarity,
$\varphi_+-\varphi_-$\,\cite{Maximon:1962zz} and
 that it is larger for small pair opening angles\,\cite{Endo:1992nq}.
Many attempts have been undertaken to increase the effective
polarization asymmetry by applying a well-chosen event selection on
the collected sample, in the hope that the sensitivity to polarization
would be increased
\cite{Kobayashi:1972sf,Depaola:1998cu,AsaiSkopik1999,Wojtsekhowski:2003gv,Adamyan:2005,Bakmaev:2007rs,deJager:2007nf}.
Actually, even though it is true that the asymmetry increases after
selection, the gain is almost entirely lost by the reduction in the
statistics of the sample, so that ultimately the gain in precision
in the polarization fraction of the incident radiation, if any, is
found to be small (section 4.1 of Ref.\,\cite{Bernard:2013jea}).

Another way out was sought by removing the W converters and using 
{\bf pure silicon converter/trackers}.
For thick wafers ($\approx 500\,\micro\meter$
\cite{TIGRE:2001,MEGA:2005,CAPSiTT:2010,Morselli:2014fua,Moiseev:2015lva,E-Astrogam:2016}),
the effective polarization asymmetry, and therefore the dilution
factor, are still low because multiple scattering remains an issue.
It's only if very thin wafers can be made, held and launched
($\approx 150\,\micro\meter$ \cite{Wu:2014tya}) that there is some hope 
of a sizeable sensitivity to polarization 
(see Figs.\,5 and\,6 of Ref.\,\cite{Wu:2014tya}, though).

Another avenue is the use of a very high-resolution dense homogeneous
active target such as an {\bf emulsion detector}\,\cite{Takahashi:2015jza}.
GRAINE has demonstrated the detection of a linear polarization signal
with pairs in a 0.8 - 2.4\,GeV $\gamma$-ray beam at SPring8, at a
$3\,\sigma$ significance level\,\cite{Ozaki:2016gvw} but their ability
to get below 100\,MeV where most of the statistics lie, remains an
issue\footnote{
Note that polarimetry with pairs has been demonstrated at high energy
(1.5 - 2.4\,GeV) but in a configuration appropriate for use on a
$\gamma$-ray beam and not for a space polarimeter, in particular since
it has an efficiency of only $0.02\%$\,\cite{deJager:2007nf}.}.
Being sensitive at low energies is critical to $\gamma$-ray
polarimetry because the polarization asymmetry increases there and,
more importantly, because the product of the conversion cross section
and of the typical $1/E^2$ flux of a cosmic source peaks at
$\approx 6\,\mega\electronvolt$
(Fig.\,2 of Ref.\,\cite{Bernard:2013jea}).

The extreme solution to the multiple scattering hurdle is the 
use of low density, that is, of {\bf gas detectors}.
The HARPO project used a time projection chamber (TPC) in which
the conversion takes place in a fast-electron-drift, 
low-diffusion argon-based gas to
perform the first demonstration of polarimetry with pairs at low
energy with an excellent dilution factor\,\cite{Gros:SPIE:2016}.
It was demonstrated that a $\gamma$-ray polarimeter can be triggered and event reconstruction performed down to the lowest $\gamma$-ray energy 
of 1.74\,MeV that the polarized $\gamma$-ray beam line BL01 of the NewSUBARU electron storage ring could provide\,\cite{Bernard:NASA2016}.

There is good prospects that the rapid degradation of the TPC gas
quality that the EGRET spark chambers have undergone and that
motivated a refill per year, is well under control with TPC
proportional gas amplifiers \cite{Hill:2013,Frotin:2015mir}.

The possibility to perform astronomy, that is, photon-to-source
assignment, on a single photon basis, down to so low an energy is
uncertain especially due to the huge angular spread induced by the
fact that the momentum of the recoiling ion cannot be measured 
(a parametrization of this contribution to the single-photon angular
resolution can be found in section 3.1.2 of
Ref.\,\cite{Bernard:2012uf}).
For GRBs though, source assignment and trigger are taken care of
globally for the burst, and the low energy $\gamma$ rays can be logged,
reconstructed and analyzed.
For pulsars, measuring the phase difference with respect to the pulsar ephemerides enables to subtract the non pulsed background.

A similar project, which uses a negative-ion slow-electron-drift gas,
 is under development\,\cite{Hunter:2013wla} 
in the hope to get a device with enough amplification gain in the gas 
later this year\,\cite{Timokhin:NASA2016}.
The low electron drift velocity (more than three orders of magnitude
lower than that in noble-gas-based mixtures) and the pile-up of stray
tracks from proton cosmic rays and from scattering of low-energy
photons might, however, make negative-ion TPCs inappropriate for use
in space.

\subsection{Kinematics}
\label{subsec:kine}

We name $E, \vec{k}$ the energy and the momentum of the incident photon.
Its direction defines the $z$ direction.
We name $E_i, \vec{p_i}$ the energy and the momentum of the
particles in the final state, with 
$i = +$ for the positron, 
$i = -$ for the electron and 
$i = r$ for the recoiling particle, that is the target ion (nuclear conversion) or electron (triplet conversion).
\iffalse
\begin{figure}[!h]
 \begin{center}
 \setlength{\unitlength}{0.47\textwidth}
 \begin{picture}(1,0.7)(0,0)
 \put(0,0){
\includegraphics[width=0.47\textwidth]{Drawing6.pdf}
}
 \put(0.27,0.5){$x$}
 \put(0.6,0.71){$y$}
 \put(0.97,0.5){$z$}
 \put(0.32,0.57){$\varphi_+$}
 \put(0.575,0.4){$\varphi_-$}
 \put(0.775,0.27){$\omega$}
 \put(0.41,0.29){$\theta_+$}
 \put(0.58,0.265){$\theta_-$}
 \put(0.3,0.3){$\vec{p_+}$}
 \put(0.59,0.2){$\vec{p_-}$}
 \put(0.3,0.02){$\vec{p_r}$}
 \put(0.03,-0.01){$\vec{k}$}
 \end{picture}
 \caption{Schema of a photon conversion.
 \label{fig:schema:angles}}
 \end{center}
\end{figure}
\else
\begin{figure}[!h]
 \begin{center}
 \setlength{\unitlength}{0.47\textwidth}
 \begin{picture}(1,0.7)(0,0)
 \put(0,0){
\includegraphics[width=0.47\textwidth]{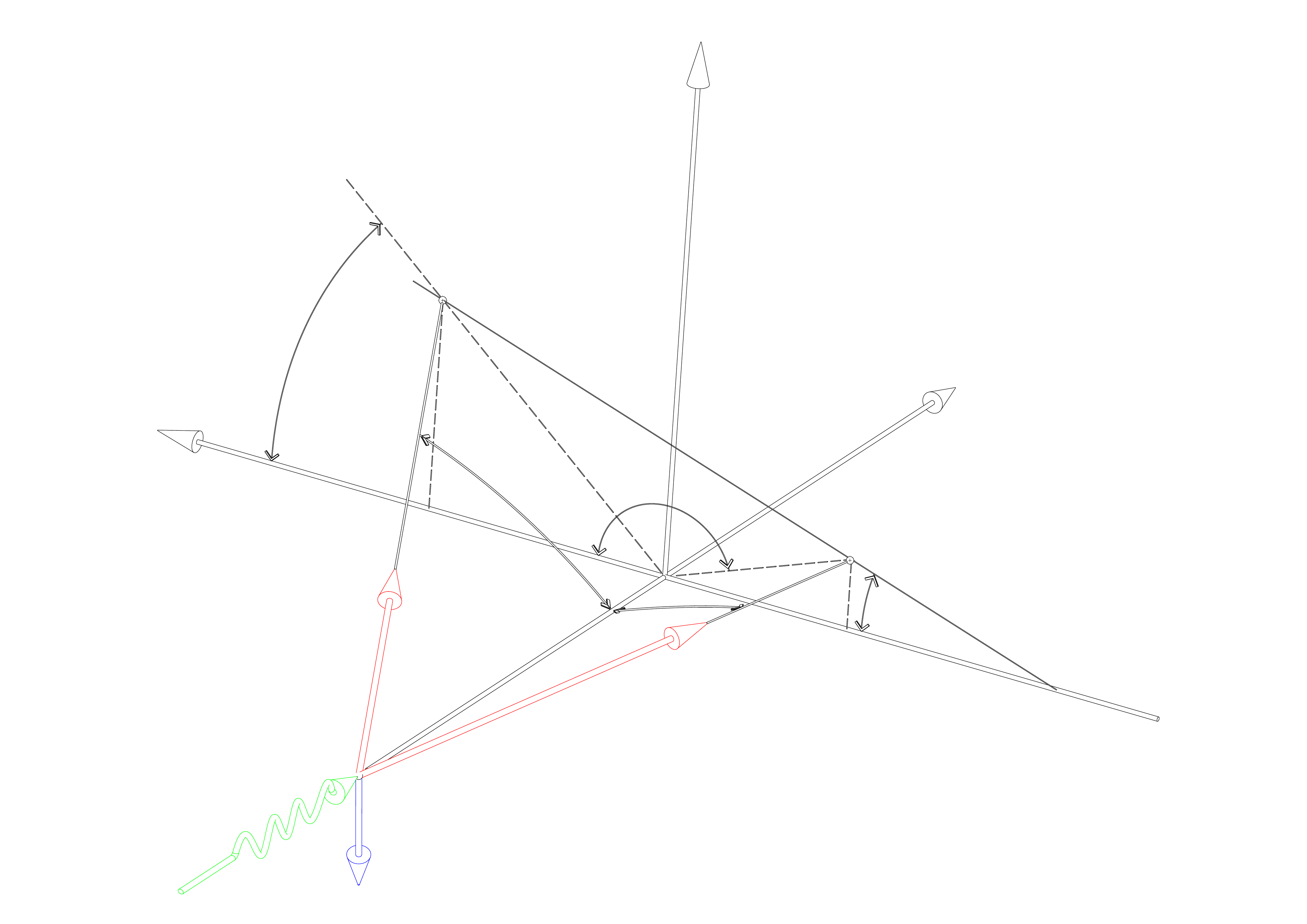}
}
 \put(0.13,0.4){$x$}
 \put(0.58,0.67){$y$}
 \put(0.7,0.44){$z$}
 \put(0.178,0.47){$\varphi_+$}
 \put(0.462,0.345){$\varphi_-$}
 \put(0.72,0.27){$\omega$}
 \put(0.39,0.24){$\theta_+$}
 \put(0.56,0.19){$\theta_-$}
 \put(0.22,0.2){\color{red}{$\vec{p_+}$}}
 \put(0.45,0.12){\color{red}{$\vec{p_-}$}}
 \put(0.32,0.02){\color{blue}{$\vec{p_r}$}}
 \put(0.08,-0.01){\color{green}{$\vec{k}$}}
 \end{picture}
 \caption{Schema of a photon conversion.
 \label{fig:schema:angles}}
 \end{center}
\end{figure}
\fi
The $e^+ e^-$ pair is denoted by the index $p$, so for example 
$ \vec{p_p} = \vec{p_+} + \vec{p_-}$.
We name $\varphi_i$ and $\theta_i$ the azimuthal and the polar angles, respectively, for $i = +, -, r, p$ (Fig.\,\ref{fig:schema:angles}).
The unitary vector that shows the direction of a particle is 
$ \vec{u_i} \equiv \vec{p_i} / p_i$ and 
the fraction of the incident photon energy that is carried away by
particle $i$ is $x_i \equiv E_i / E$.

\subsection{Paper layout}
\label{subsec:paper:layout}

In this paper, we examine in detail the energy variation of the
polarization asymmetry of $\gamma$-ray nuclear conversion to $e^+e^-$ pairs,
using the polarized, fully 5-dimensional (5D), exact down to threshold
event generator that was documented in Ref.\,\cite{Bernard:2013jea}.
\begin{enumerate}
\item It has been known since the last century that the asymmetry
 increases at low $\gamma$-ray energies.
The high-energy (HE) asymptotic variation was obtained for triplet
conversion\,\cite{Boldyshev:1972va} and as the same two Feynman diagrams
dominate nuclear conversion and, at high energy, triplet conversion,
the expression is also asymptotic for nuclear conversion.
{\bf We examine how the simulated value tends asymptotically to the expression of Ref.\,\cite{Boldyshev:1972va}.}

\item From the Bethe-Heitler (BH) differential cross section, 
{\bf we determine the value of the polarization asymmetry $A$ at threshold energy};
{\bf We examine how the exact value tends to the limit at threshold}.

\item 
Various expressions have been used in the past as ``the azimuthal
angle'' that references the azimuthal orientation of the final state
with respect to the direction of polarization of the incident photon:
\begin{itemize}
\item
The analytical expression\,\cite{May1951,Heitler1954} 
(see also eq.\,\ref{eq:pair:pol})
used the bisector angle $\phi \equiv (\varphi_+ + \varphi_-)/2$ of the
lepton pair.
\item
Later works used the azimuthal angle of the recoiling particle,
$\varphi_{r}$, especially for triplet conversion for which this angle
is directly measurable
\cite{Boldyshev:1994bs,Depaola:2009zz,Iparraguirre:2011zz}.
The azimuthal angle of the pair, that is, of the momentum of the pair,
$\varphi_{p}$, is equivalent to $\varphi_{r}$, as the pair and the
recoiling target are flying back-to-back in the center-of-mass (CMS)
system, $\varphi_{p} = \varphi_{r}\pm \pi$.
\item 
Experimentalists, 
\cite{deJager:2007nf,Bernard:2013jea,Takahashi:2015jza,Gros:SPIE:2016}
 following Wojtsekhowski\,\cite{CLAS-98-018,Wojtsekhowski:2003gv}
use the pair plane azimuthal angle $\omega$ (Fig.\,\ref{fig:schema:angles}):
\begin{equation}
\omega = \arctan\left(\frac{u_{-z}u_{+x}-u_{+z}u_{-x}}{u_{-z}u_{+y}-u_{+z}u_{-y}}\right) .
\label{eq:omega:1}
\end{equation}
\end{itemize}

Polarimetry of the $\gamma$ emission of a cosmic source can be
performed only after astronomy, that is, the photon-to-source
assignment, has been performed, be it on a single photon basis (steady
sources) or on a burst population basis (GRB).
Therefore a putative photon direction can always be defined 
and the measurement of $\phi$, of $\omega$ and of $\varphi_{r}$ (for
triplet) are straight-forward, while the measurement of $\varphi_{p}$
requires, in addition, a measurement of the magnitude of the momenta
of the electron and of the positron.
{\bf We examine the performance of these possible definitions of the
 azimuthal angle with the aim to maximize the obtained value of the
 polarization asymmetry} and,
therefore,
{\bf of the precision of the measurement of $P$}.
\end{enumerate}

\section{Bethe--Heitler polarization asymmetry at threshold}
\label{sec:Bethe-Heitler}

As the original work by Bethe--Heitler neglected the Feynman diagrams
for which the incident photon has its vertex with the target particle
(Feynman diagrams (c) and (d) of Fig.\,1 of
Ref.\,\cite{Bernard:2013jea}), we name here Bethe--Heitler (BH) the
calculations performed under this hypothesis.
Even though 
\begin{itemize}
\item the ion mass is much larger than the electron mass,
\item in the case of triplet conversion there are two electrons in the final
state, so that there is a set of 4 additional ``exchange'' diagrams,
\end{itemize}
the two same ``Borsellino'' diagrams that dominate the
differential cross section for nuclear conversion also dominate the
triplet conversion asymptotically at high energy.


The full 5D, unpolarized, differential cross-section was calculated by
Bethe and Heitler\,\cite{Bethe-Heitler,Heitler1954}:
\begin{eqnarray} 
\label{eq:pair:unpol}
\dd \sigma & = & \frac{- \alpha Z^2 r_0^2 m^2}{(2 \pi)^2 E ^3} \dd E_+ \dd \Omega_+ \dd \Omega_- \frac{|p_-||p_+|}{|\vec{q}|^4} 
\\
 & & 
\left[ 
\left(
\frac{p_+ \sin{\theta_+}}{E_+ - p_+ \cos{\theta_+}}
\right)^2 (4 E_-^2 - q^2)
+
\left(
\frac{p_- \sin{\theta_-}}{E_- - p_- \cos{\theta_-}}
\right)^2 (4 E_+^2 - q^2)
+
\right.
 \nonumber 
\\
 & & 
 \left. 
\frac{2p_+ p_- \sin{\theta_+} \sin{\theta_-} \cos{(\varphi_+-\varphi_-)}}{(E_- - p_- \cos{\theta_-})(E_+ - p_+ \cos{\theta_+})}
(4 E_+ E_- + q^2 -2 E ^2) 
\right. 
\nonumber
\\ 
 & & 
\left.
- 2 E ^2
\frac{(p_+ \sin{\theta_+})^2 + (p_- \sin{\theta_-})^2}
{(E_+ - p_+ \cos{\theta_+})(E_- - p_- \cos{\theta_-})}
\right] .
 \nonumber 
\end{eqnarray}

The azimuthal information that takes part explicitly in the expression
of the unpolarized differential cross section comes by the factor
$\cos{(\varphi_+-\varphi_-)}$ that expresses its dependence on the
acoplanarity of the two leptons\footnote{The $1/q^4$ factor obviously
 contributes to keep the leptons close to back-to-back.}.

Berlin and Madansky obtained the expression for a linearly polarized
photon\,\cite{BerlinMadansky1950}, that was later expressed in
Bethe-Heitler notation by May\,\cite{May1951}, and after a correction by
a factor of 2 as indicated by\,\cite{jau}, is:
\begin{eqnarray}
\dd\sigma & = & \frac{-2 \alpha Z^2 r_0^2 m^2}{(2 \pi)^2 E ^3} \dd E_+ \dd \Omega_+ \dd \Omega_- \frac{|p_-||p_+|}{|\vec{q}|^4} 
\times \\
 & & \left[ 
\left(
2 E_+ \frac{p_- \sin{\theta_-} \cos{\varphi_-}}{E_- - p_- \cos{\theta_-}} + 
2 E_- \frac{p_+ \sin{\theta_+} \cos{\varphi_+}}{E_+ - p_+ \cos{\theta_+}}
\right)^2
 \right. \nonumber 
\\
 & & -q^2 
\left. 
\left(
\frac{p_- \sin{\theta_-} \cos{\varphi_-}}{E_- - p_- \cos{\theta_-}} - 
\frac{p_+ \sin{\theta_+} \cos{\varphi_+}}{E_+ - p_+ \cos{\theta_+}}
\right)^2
 \right. \nonumber 
\\
 & & -E^2 
\left. 
\frac{(p_+ \sin{\theta_+})^2+(p_- \sin{\theta_-})^2+2p_+ p_- \sin{\theta_+} \sin{\theta_-} \cos{(\varphi_+-\varphi_-)}}{(E_- - p_- \cos{\theta_-})(E_+ - p_+ \cos{\theta_+})}
\right] .
 \nonumber 
\label{eq:pair:pol}
\end{eqnarray}
with
\begin{eqnarray}
|\vec{q}|^2 = |\vec{p_+} + \vec{p_-} - \vec{k} |^2 .
\label{eq:def:q2}
\end{eqnarray}

Polarimetrists re-cast the above expressions for a partially polarized
beam with linear polarization fraction $P$ as:
\begin{eqnarray}
\dd \sigma = \Phi(X_{u} + P \times X_{p})
\dd E_+ \dd \Omega_+ \dd \Omega_- ,
\label{eq:dsig:short}
\end{eqnarray}
 with:
\begin{eqnarray}
 \Phi = \frac{- \alpha Z^2 r_0^2 m^2}{(2 \pi)^2 E ^3} \frac{|p_-||p_+|}{|\vec{q}|^4} ,
\label{eq:pair:BigPhi}
\end{eqnarray}
 and
\begin{eqnarray}
X_{u}& = &
\left[ 
\left(
\frac{p_+ \sin{\theta_+}}{E_+ - p_+ \cos{\theta_+}}
\right)^2 (4 E_-^2 - q^2)
+
\right.
\nonumber 
\\
& & 
\left.
\left(
\frac{p_- \sin{\theta_-}}{E_- - p_- \cos{\theta_-}}
\right)^2 (4 E_+^2 - q^2)
+
\right.
\nonumber 
\\
 & & 
 \left. 
\frac{2p_+ p_- \sin{\theta_+} \sin{\theta_-} \cos{(\varphi_+-\varphi_-)}}{(E_- - p_- \cos{\theta_-})(E_+ - p_+ \cos{\theta_+})}
(4 E_+ E_- + q^2 -2 E ^2) 
- 
\right.
\nonumber 
\\
& & 
\left.
2 E ^2
\frac{(p_+ \sin{\theta_+})^2 + (p_- \sin{\theta_-})^2}
{(E_+ - p_+ \cos{\theta_+})(E_- - p_- \cos{\theta_-})}
\right] ,
\label{eq:pair:unpol:X}
\end{eqnarray}
\begin{eqnarray}
X_{p} & = & 
\cos{2\varphi_-} 
(4 E_+^2 -q^2)
 \left(
\frac{p_- \sin{\theta_-}}{E_- - p_- \cos{\theta_-}} \right)^2
+
\nonumber 
\\
& & 
\cos{2\varphi_+} 
(4 E_-^2 -q^2)
 \left(
 \frac{p_+ \sin{\theta_+}}{E_+ - p_+ \cos{\theta_+}}
\right)^2 
 \nonumber 
\\
 & & 
+2 \cos{(\varphi_++\varphi_-)} (4 E_+ E_- +q^2)
\frac{p_- \sin{\theta_-} p_+ \sin{\theta_+}}{(E_- - p_- \cos{\theta_-})(E_+ - p_+ \cos{\theta_+})} .
\label{eq:pair:pol:X}
\end{eqnarray}
 
Close to threshold
$E \approx 2 m \gg p_{\pm}$, ~ 
$q \approx E$, ~
$E_{\pm} \approx m$, ~ 
$E_{\pm} - p_{\pm} \cos{\theta_{\pm}} \approx m$ 
\cite{Boyarkin} and the differential cross section becomes:
\begin{eqnarray}
\dd \sigma & = & 
 \frac{\alpha Z^2 r_0^2}{\pi^2 64 m^5} p_- p_+
\dd E_+ 
\sin{\theta_+} \dd \theta_+ \dd \varphi_+ 
\sin{\theta_-} \dd \theta_- \dd \varphi_- 
\nonumber \\
& & 
\left(
(p_+ \sin{\theta_+})^2 + (p_- \sin{\theta_-})^2 
 - 2 P 
 \cos{(\varphi_++\varphi_-)} p_- \sin{\theta_-} p_+ \sin{\theta_+}
 \right) ,
~ ~ ~ ~ ~ ~
\label{eq:dsig:long:LE}
\end{eqnarray}

which, after integration over the electron and positron polar angles, becomes:
\begin{eqnarray}
\dd \sigma & = & 
 \frac{\alpha Z^2 r_0^2}{\pi^2 64 m^5}
\dd E_+ 
\dd \varphi_+ 
\dd \varphi_- 
\left((p_+^2 + p_-^2) p_- p_+ \gfrac{8}{3}
 - P 
 \cos{(\varphi_++\varphi_-)} (p_- p_+)^2 \gfrac{\pi^2}{2} 
 \right) .
~ ~ ~ ~ ~ ~
\label{eq:dsig:long:LE:ter}
\end{eqnarray}

Integrating over $E_+$ and remembering the expression for the bisector
angle of the lepton pair, $\phi \equiv (\varphi_+ + \varphi_-)/2$, we
finally obtain:
\begin{eqnarray}
\dd \sigma & = & 
 \frac{\alpha Z^2 r_0^2}{3 \pi 16 m^3}
\dd \varphi_+ 
\dd \varphi_- 
\left(E - 2 m \right)^3
\left(1 - P \gfrac{\pi}{4} \cos{2\phi} \right) .
\label{eq:dsig:long:LE:quint}
\end{eqnarray}
\begin{itemize}
\item
Further integration over the lepton azimuthal angles yields the
well-known low-energy asymptote for the Bethe-Heitler total cross section 
\begin{eqnarray}
 \sigma & = & 
 \alpha Z^2 r_0^2
\left(\gfrac{E - 2 m}{m} \right)^3 \gfrac{\pi}{12} .
\label{eq:sig:LE:leretour}
\end{eqnarray}
\item
From eq.\,\ref{eq:dsig:long:LE:quint} we obtain the 
 low-energy asymptote for the $\gamma$-conversion polarization asymmetry 
\begin{eqnarray}
A = \gfrac{\pi}{4} .
\label{eq:sig:LE:final}
\end{eqnarray}
\item The high-energy asymptotic expression of the singly differential
 cross section
 is\,\cite{Boldyshev:1972va}
\begin{eqnarray}
2\pi \gfrac{\dd \sigma}{\dd \phi} \propto
\alpha r_0^2
 \left[
 \left(\gfrac{28}{9}\ln{2 E} - \gfrac{218}{27} \right)
- P \cos{2\phi} 
 \left(\gfrac{4}{9}\ln{2 E} - \gfrac{20}{28} \right)
 \right],
\end{eqnarray}
from which
\begin{eqnarray}
A \approx \gfrac
{\gfrac{4}{9}\ln{2 E} - \gfrac{20}{28}}
{\gfrac{28}{9}\ln{2 E} - \gfrac{218}{27}} ,
\label{eq:sig:HE}
\end{eqnarray}
which provides the high-energy asymptotic value of the
polarization asymmetry $A = 1/7 \approx 14.3\%$.
\end{itemize}

\section{The pair conversion event generator}
\label{sec:event:generator}

We use here a polarized event generator based on the
SPRING\,\cite{Kawabata:1995th} event generator, the 5D differential
cross section being either the Bethe-Heitler analytical
expression, Eqs.\,\ref{eq:pair:unpol}--\ref{eq:pair:pol:X}, 
(that includes only the two dominant Feynman diagrams)
or a full diagram computation using the HELAS amplitude
calculator\,\cite{Murayama:1992gi}.
In the present section we briefly summarize the documentation and the validation cross-checks of this generator that were published in Ref.\,\cite{Bernard:2013jea}.

The two possible computations of the differential cross section are
exact down to threshold, which means that no high-energy approximation
is made.
The final state is determined by five variables that are chosen to
be the polar and azimuthal angles of the electron and of the positron
and the fraction of the incident photon energy that is carried away by the positron, namely 
$\theta_-$, 
$\varphi_-$, 
$\theta_+$, 
$\varphi_+$ and
$x_+$. 
Energy-momentum conservation in the conversion is strictly enforced
both when using differential cross sections from HELAS or from
Bethe-Heitler, even though the early uses of the Bethe-Heitler
formalism assumed that the energy carried away by the target was
negligible.
Event generation was qualified by confrontation with analytical
results on 1D projections published in the past
(Figs.\,4--6 and (supplementary data) Figs.\,28 and 29
of Ref.\,\cite{Bernard:2013jea}).

Screening of the electric field of the nucleus by the atomic electrons
plays a role at high energies only (\cite{Bernard:2013jea} and
references therein) and is therefore not simulated in the present work.
Note also that the Coulomb corrections due to the electrostatic
interactions between the charged particles in the final state are not
taken into account in this generator.

\section{Polarization asymmetry measurement}
\label{sec:measurement}

To measure a polarization fraction, whatever the definition of the
azimuthal angle $\varphi$, we can either fit its distribution
(eq.\,\ref{eq:modulation}) or use a moments method.
The average $\langle w \rangle$ of a function named weight
$w(\varphi)$ of a variable $\varphi$ is computed over the event
sample.
If $\langle w \rangle$ depends on the value of the parameter to be
measured (here $A \times P$), $A \times P$ can be obtained from the
value of $\langle w \rangle$.

Here we extend the result of Ref.\,\cite{Bernard:2013jea} to the case
for which the polarization angle $\varphi_0$ is not known:
\begin{eqnarray}
A&=&2\sqrt{\langle \cos{2\varphi}\rangle^2+ \langle \sin{2\varphi}\rangle^2} ,
\\
\varphi_0&=&\frac{1}{2}\arctan\left(\frac{\langle \sin{2\varphi}\rangle}{\langle \cos{2\varphi}\rangle}\right) ,
\label{eq:fit}
\end{eqnarray}
The interest is that the weight can be chosen to be an optimal
estimator for $A \times P$ (\cite{Bernard:2013jea} and references
therein).
In the case of a weight built on a single variable (here $\varphi$)
the moments method is therefore equivalent to a likelihood fit of
eq.\,\ref{eq:modulation}
(that we confirm by comparing the results of the two methods).

When instead the weight is built on the whole set of variables that
describe the final state, the moments method is not only optimal but
also much simpler than an $n$-dimensional likelihood fit.
In the following we also present the results obtained with such a weight
(eq.\,11 of Ref.\,\cite{Bernard:2013jea}), that we name here ``5D''.
Moments methods provide a simple and robust estimator, which is very
welcome especially in the case of a highly-dimensional final state for
which efficiency correction is made difficult by correlations between
variables and/or the multidimensional distribution of the background(s)
is unknown : the analyst just needs to simulate the moments related to
the efficiency and/or to measure the moments related to the background,
for correction, without having to accurately parametrize the $n$-dimensional shapes of the efficiency and/or of the background\,\cite{Aubert:2004cp}.
\begin{figure}[ht] 
\includegraphics[width=0.32\linewidth]{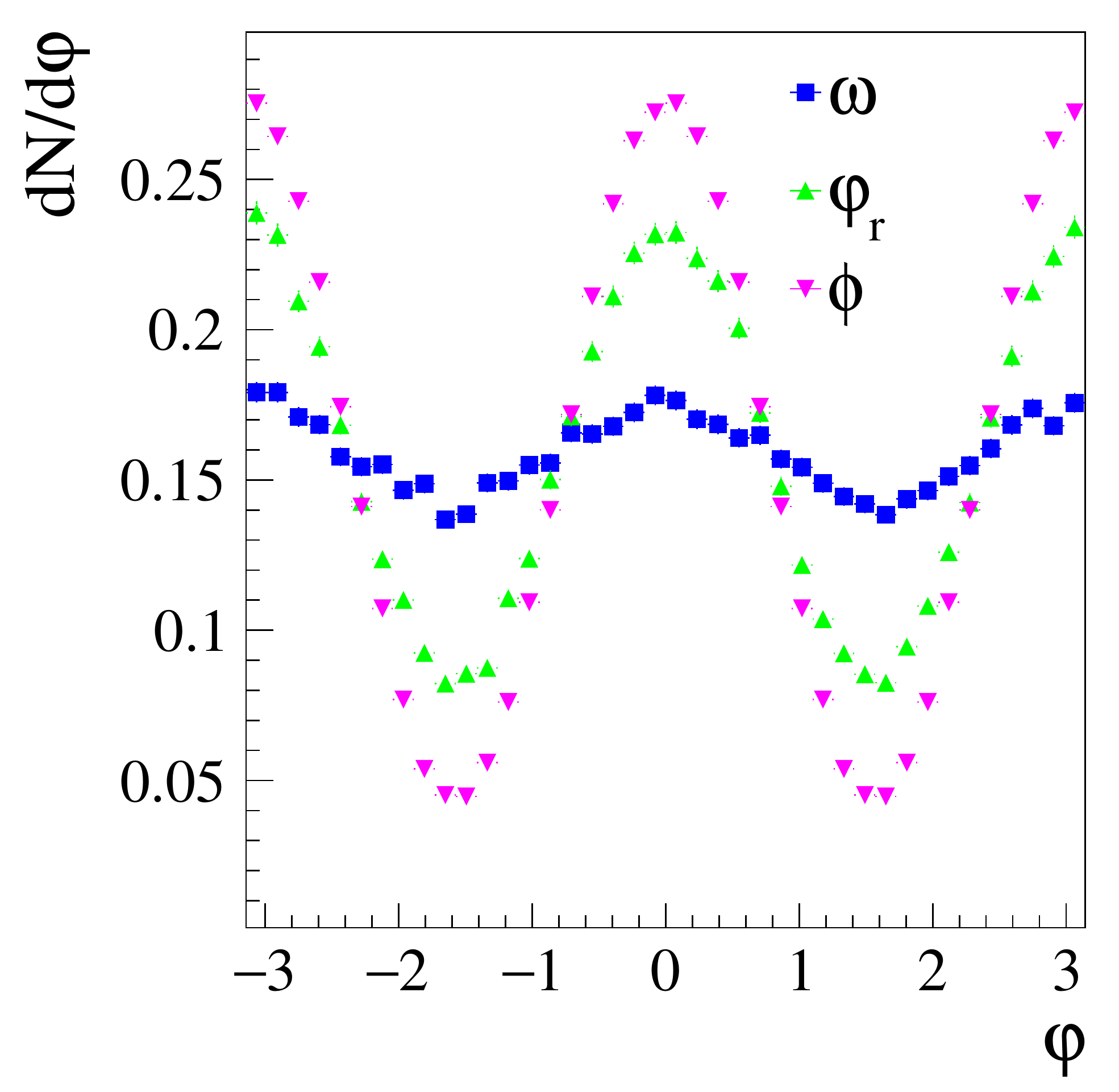}
\hfill
\includegraphics[width=0.32\linewidth]{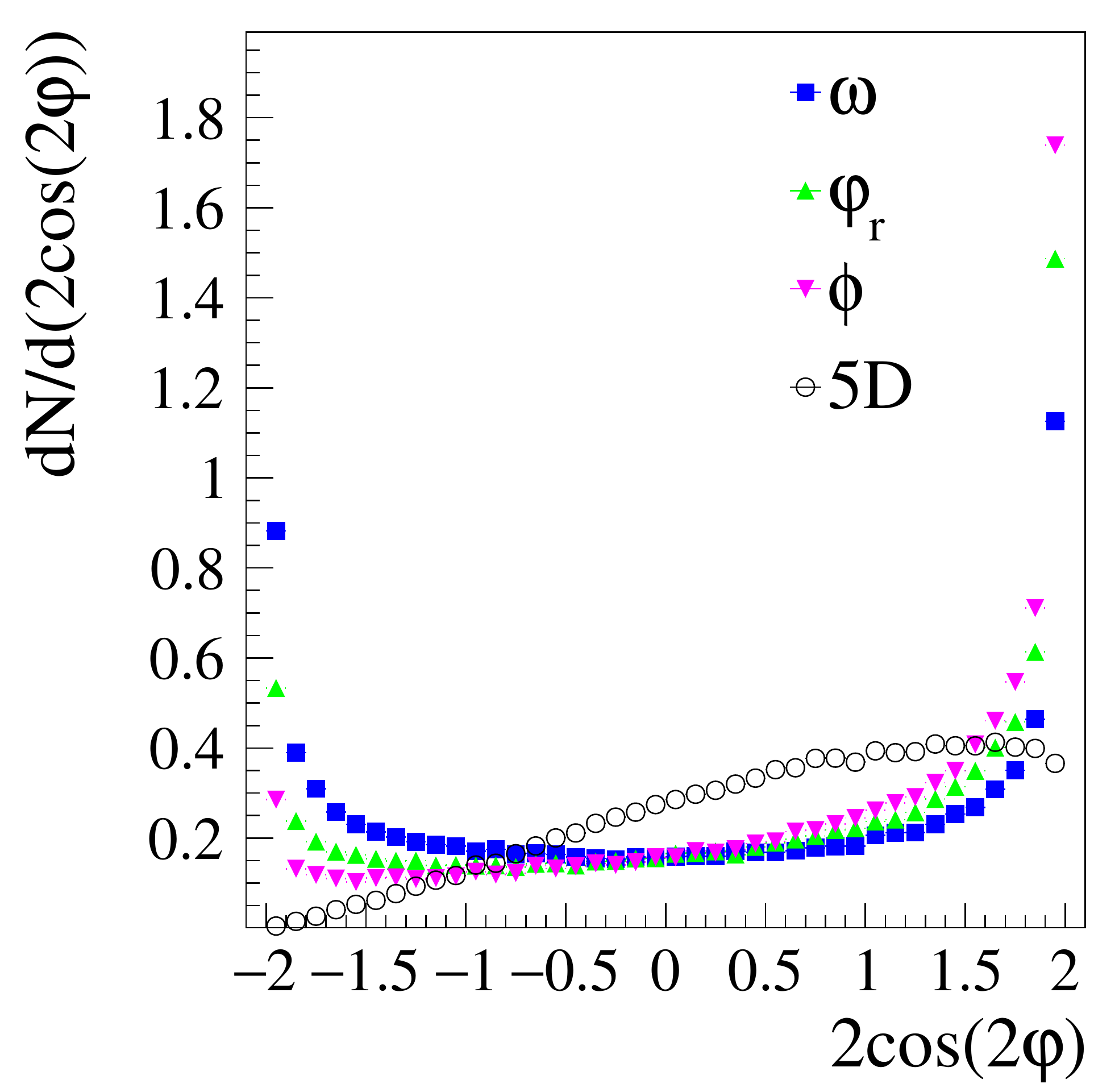}
\hfill
\includegraphics[width=0.32\linewidth]{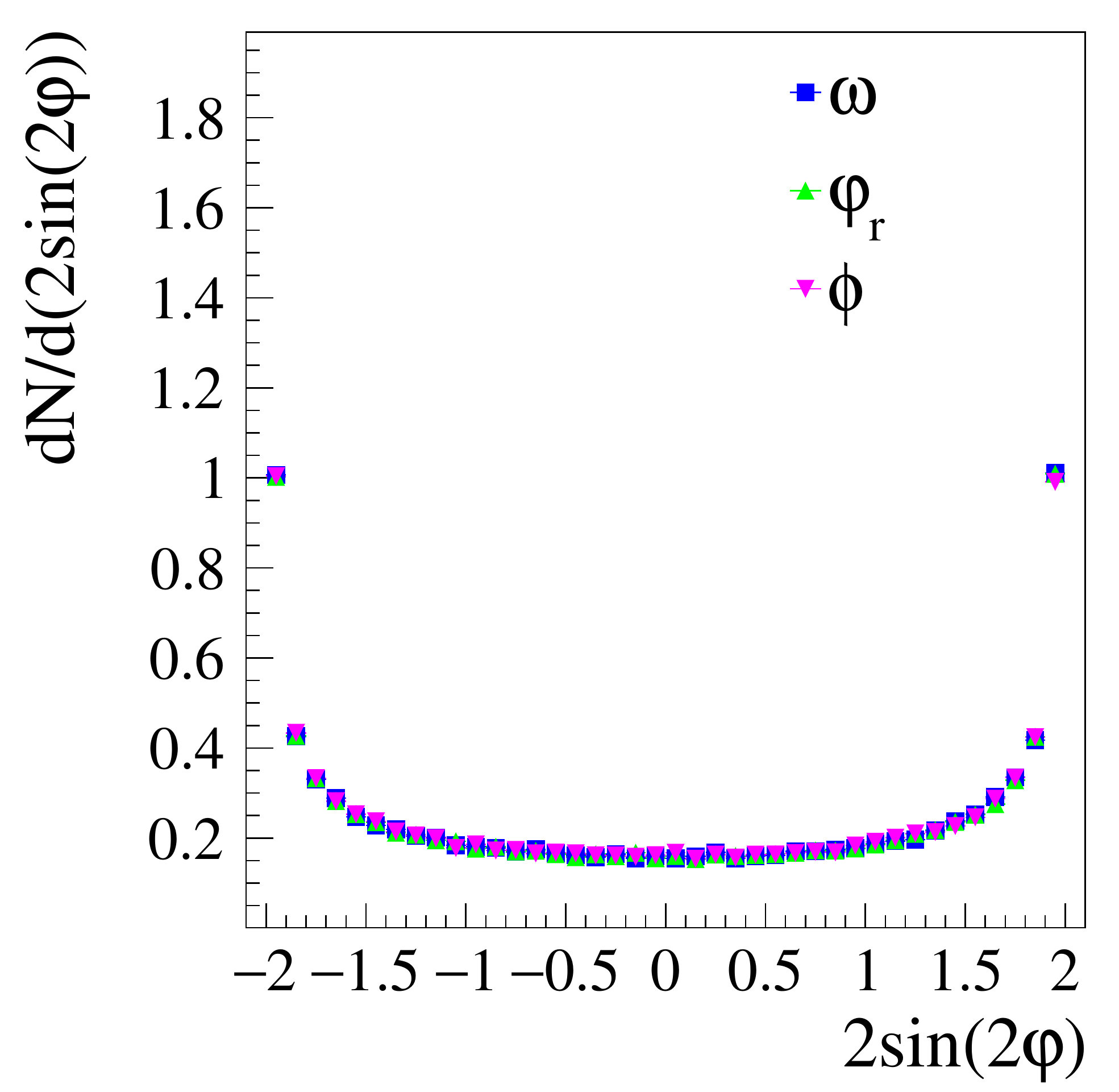}
\hfill
\put(-290,25){$P=1$}

\includegraphics[width=0.32\linewidth]{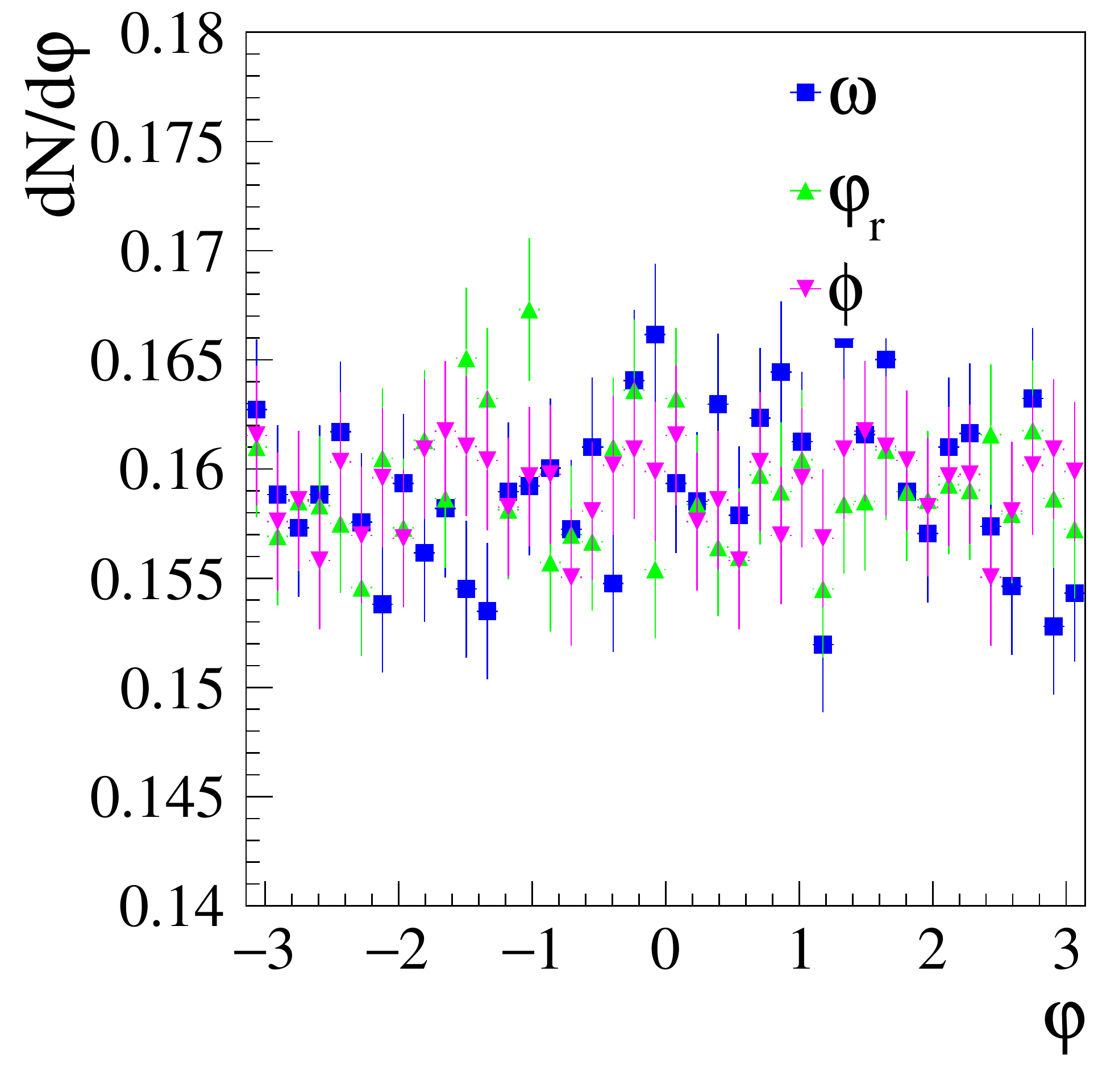}
\hfill
\includegraphics[width=0.32\linewidth]{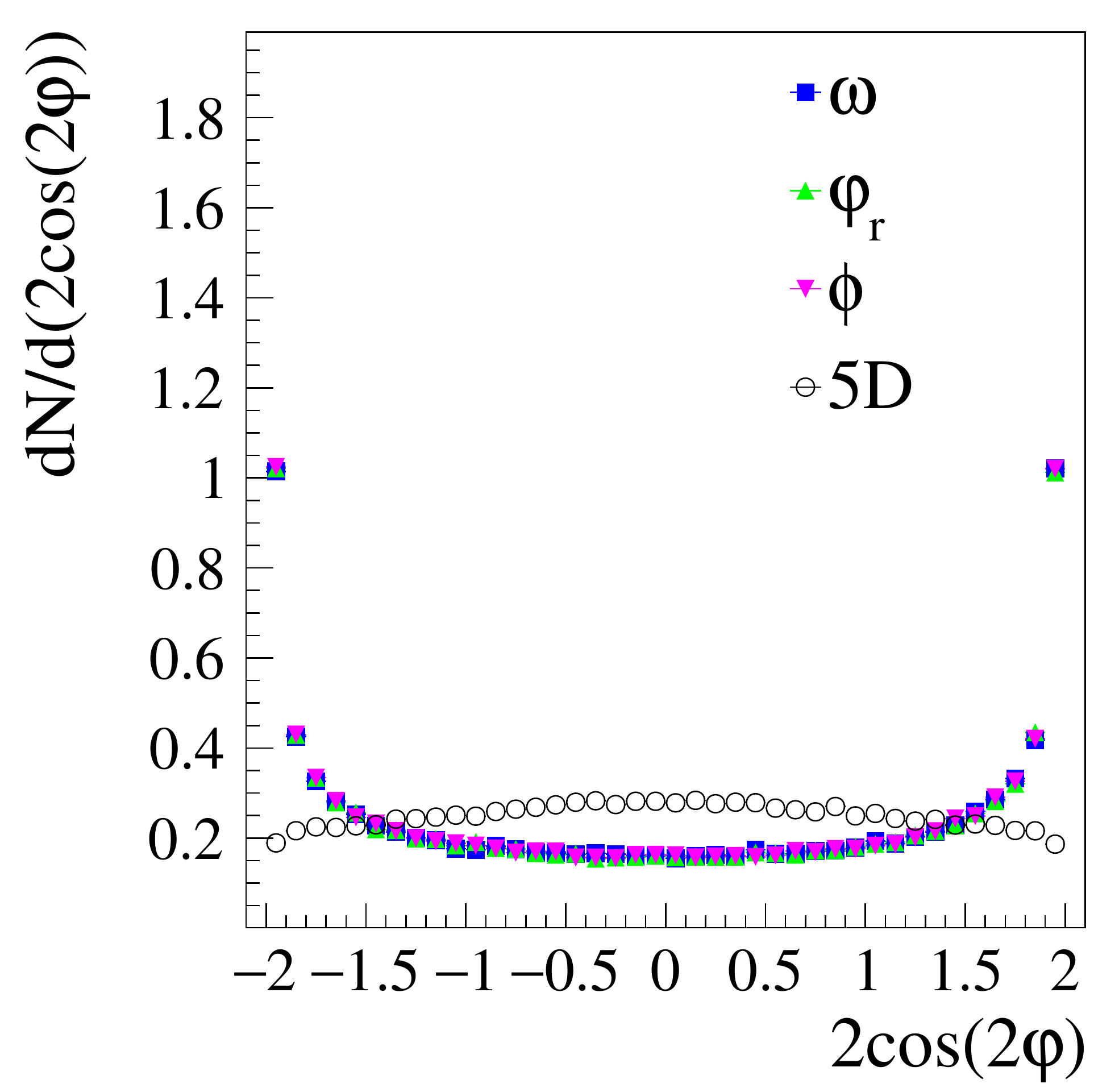}
\hfill
\includegraphics[width=0.32\linewidth]{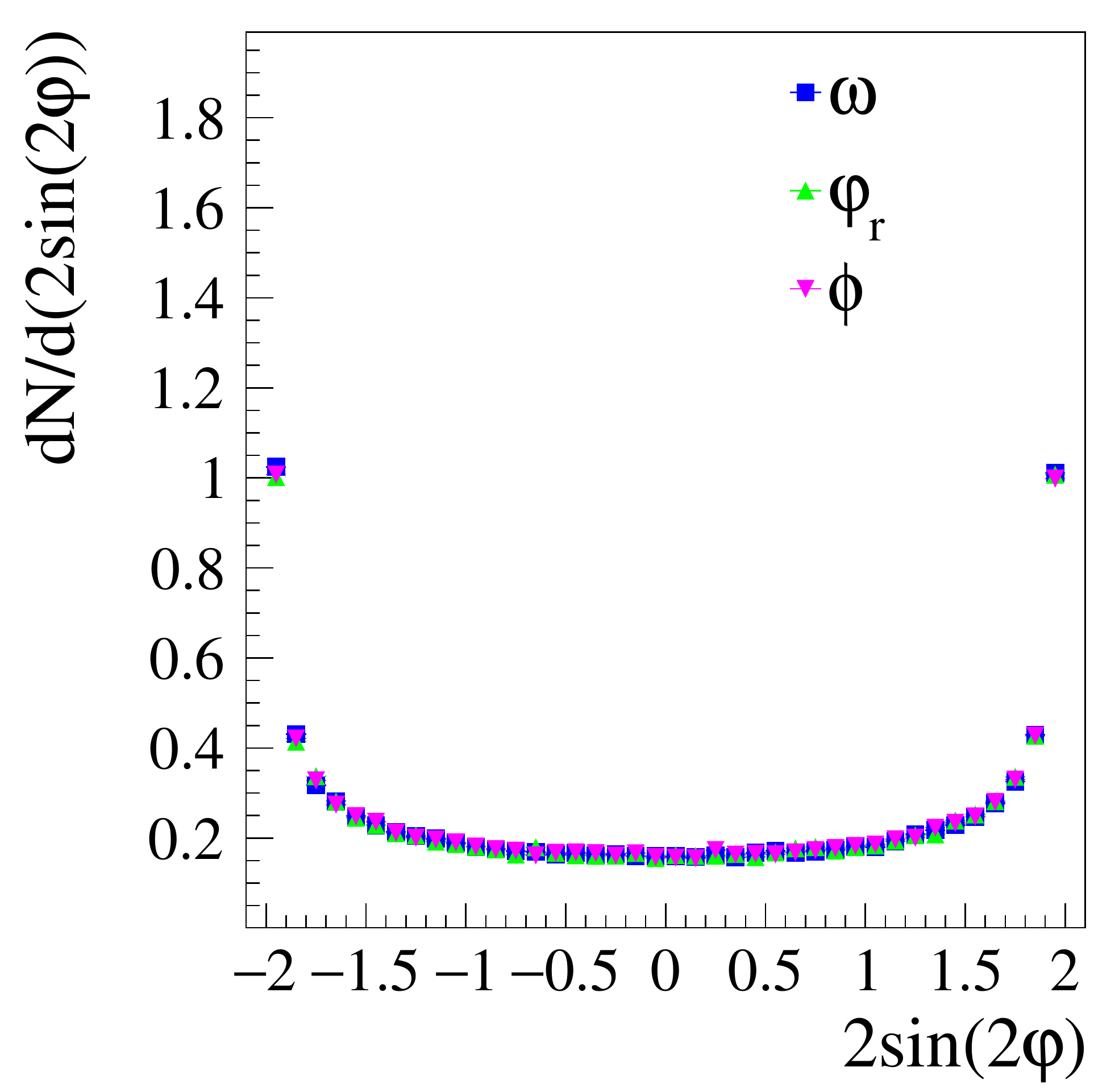}
\put(-290,25){$P=0$}
 \caption{Distributions of
 $\varphi$ (left), of $2\cos{2\varphi}$ (center),
and of $2\sin{2\varphi}$ (right) for various
definitions of angle $\varphi$ 
(squares: $\omega$,
upward triangles: $\varphi_r$,
downward triangles: $\phi$), 
for samples of $10^5$ events generated with an incident photon energy
of $E=1.2\,\mega\electronvolt$.
On the $\cos$-based plot, the distribution of the 5D weight is also shown (circles).
Upper row: $P=1$, bottom row: $P=0$. 
\label{fig:angle:distributions}
 }
\end{figure}

We have generated samples of $N=10^5$ events each, for photon energies
ranging from 1.1\,MeV to 1\,GeV, a polarization angle of
$\varphi_0 = 0$ and with either $P=1$ or $P=0$.
Figure \ref{fig:angle:distributions} presents the $\varphi$
distributions (left), the $2\cos{2\varphi}$ distributions (center),
and the $2\sin{2\varphi}$ distributions (right) for the various
definitions of angle $\varphi$ (upper row: $P=1$, bottom row: $P=0$).
The polarigrams show a clear modulation of the azimuthal angle for the
polarized event sample, for which the amplitude varies depending on
the expression that is used to estimate this angle.
The distribution of the weight $2\cos{2\varphi}$ shows a strong 
asymmetry for $P=1$, as is expected for a radiation polarized along $x$, 
none for $P=0$,
and the distribution of the weight $2\sin{2\varphi}$ doesn't show any
asymmetry as can be expected for these samples generated with
$\varphi_0=0$.

\begin{figure}[ht] 
 \includegraphics[width=0.45\linewidth]{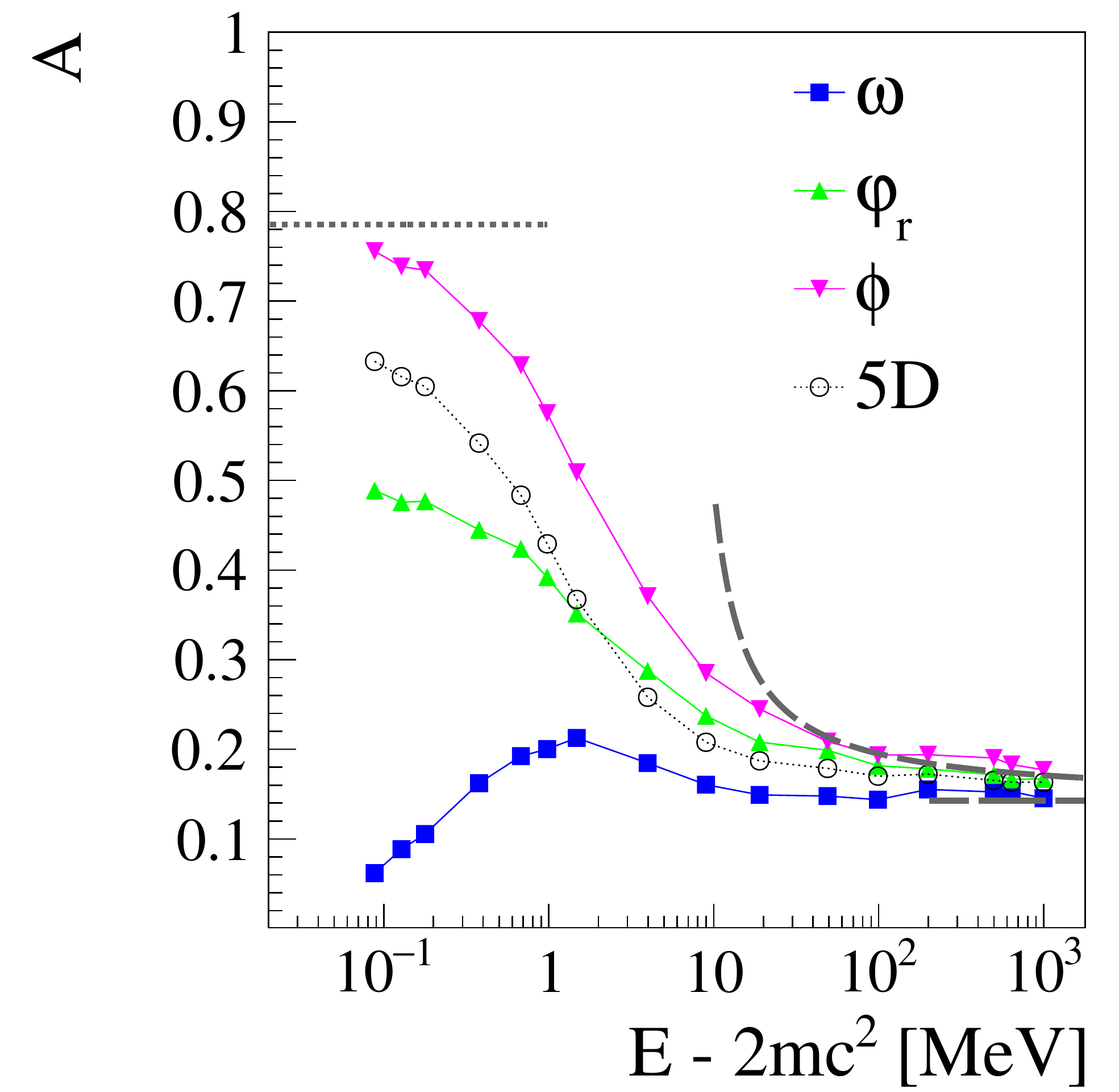}
\put(-110,130){{\bf (a)}}
\put(-153,119.5){$\gfrac{\pi}{4}$ ...}
\put(-7,38){$-$}
\put(4,38){$\gfrac{1}{7}$}
\hfill
 \includegraphics[width=0.45\linewidth]{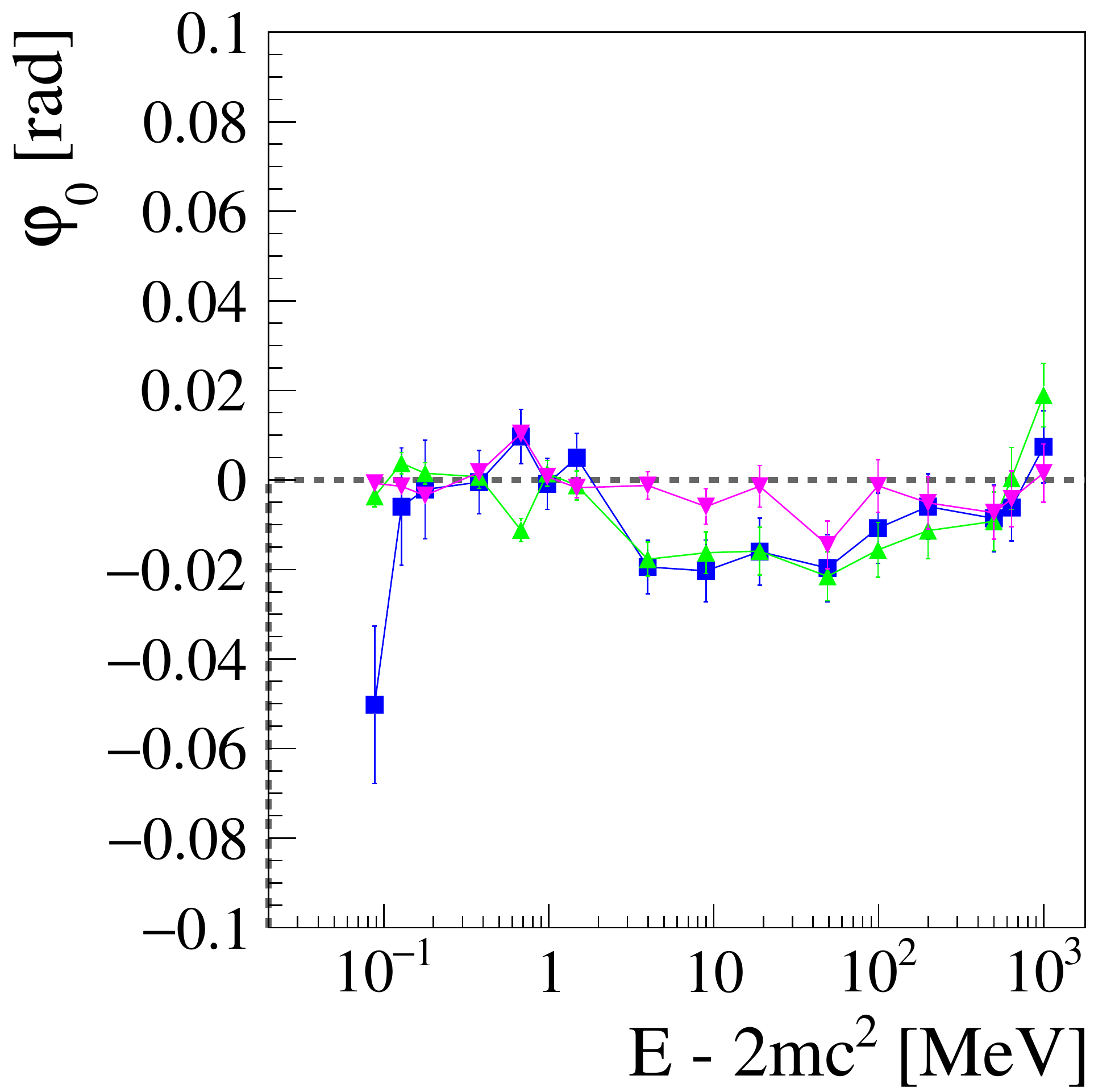}
\put(-110,130){{\bf (b)}}

 \includegraphics[width=0.45\linewidth]{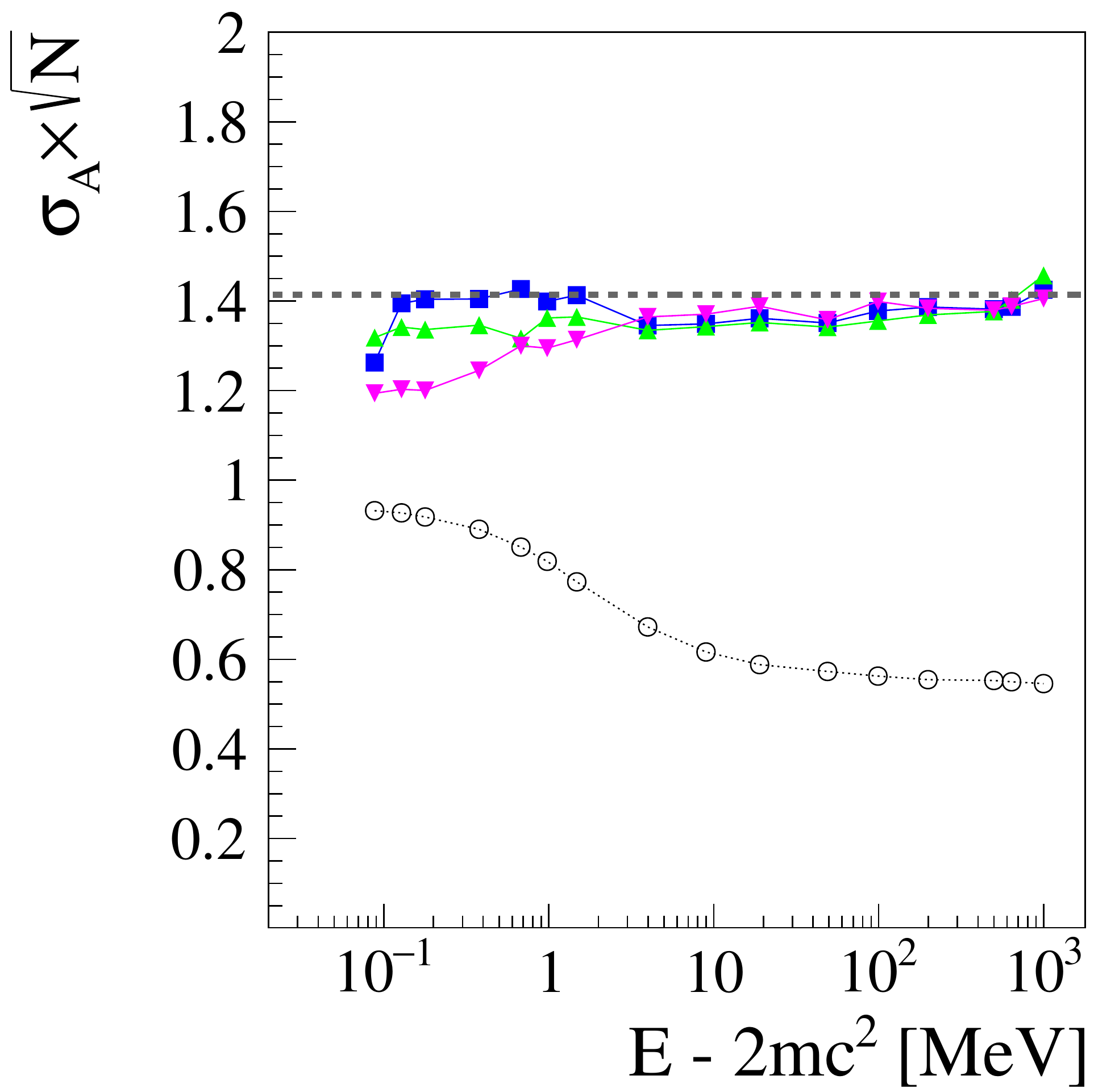}
\put(-110,130){{\bf (c)}}
\put(-7,108){$-$}
\put(4,108){$\sqrt{2}$}
\hfill
 \includegraphics[width=0.45\linewidth]{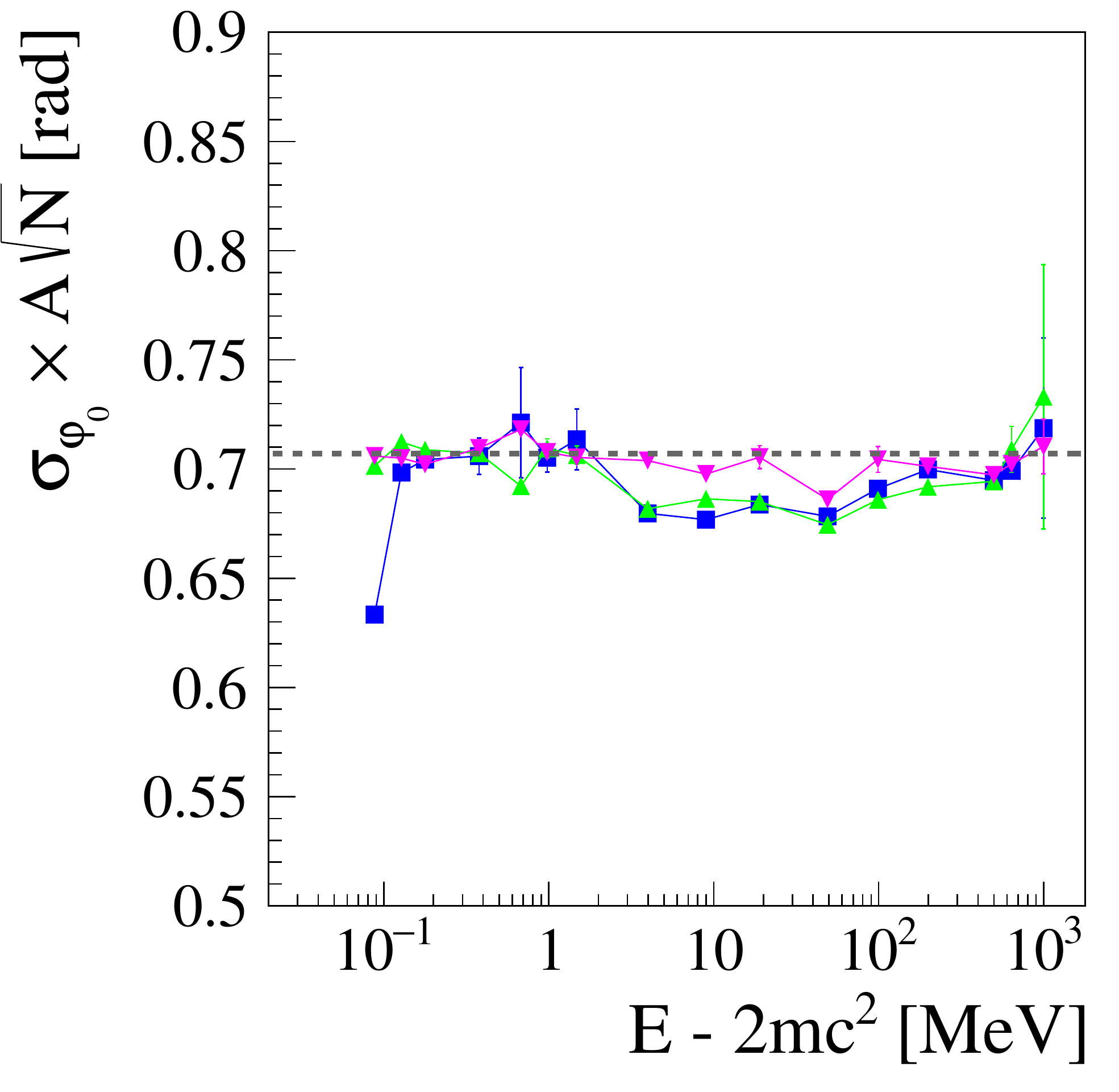}
\put(-110,130){{\bf (d)}}
\put(-7,86){$-$}
\put(4,86){$\gfrac{1}{\sqrt{2}}$}

 \includegraphics[width=0.45\linewidth]{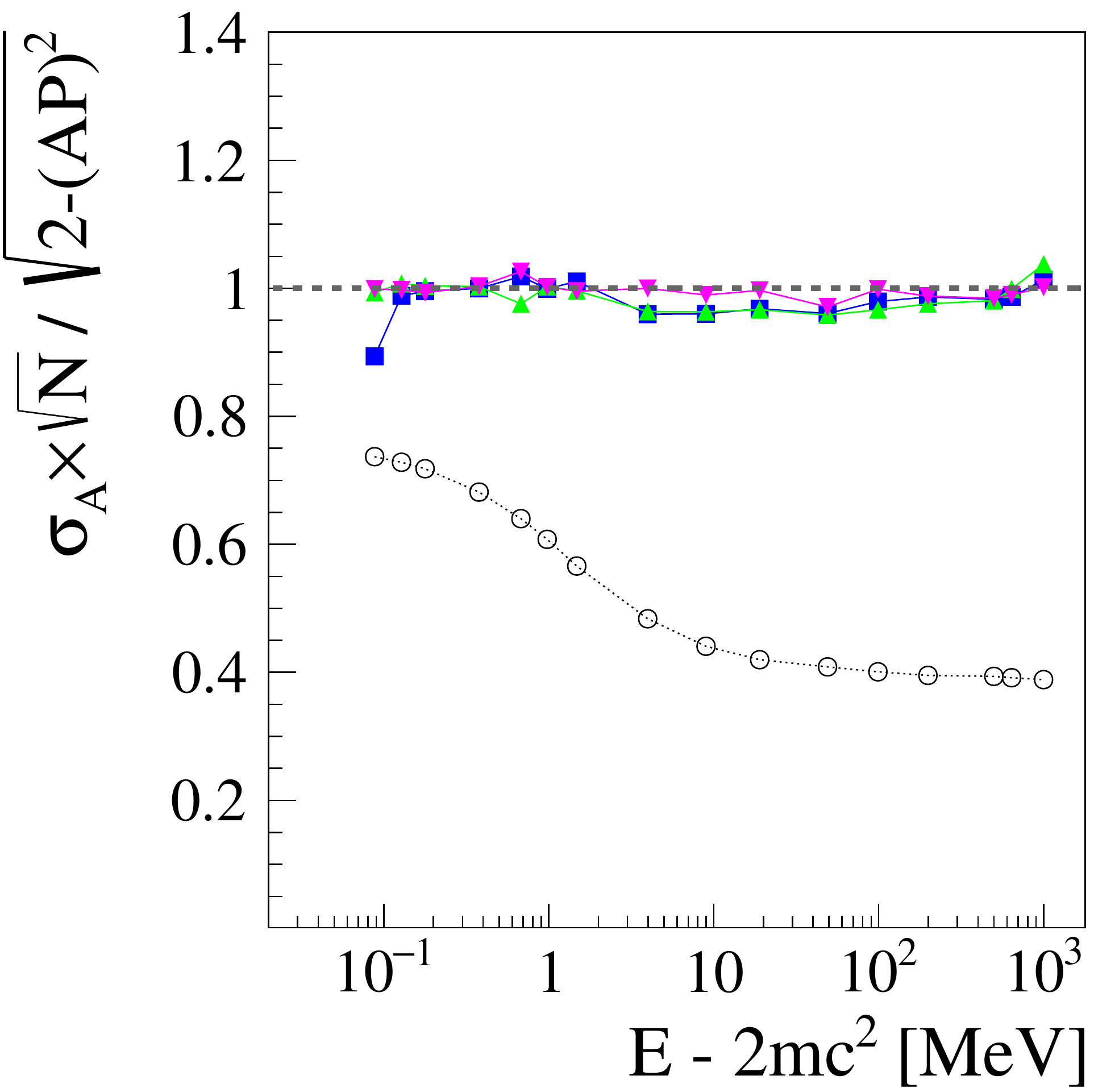}
\put(-110,130){{\bf (e)}}
 \caption{
Amplitude $A$ (left) and phase $\varphi_0$ (right) of the modulation of the distribution for various definitions of the azimuthal angle
(squares: $\omega$,
upward triangles: $\varphi_r$,
downward triangles: $\phi$.
In addition, the performance of the 5D estimator is shown (circles).
Measured value (top row), and uncertainty (middle row).
The dotted lines show: 
{\bf (a)}
the asymptotic values of $A = \pi/4$ at low energy;
{\bf (c)}
the approximate value of the uncertainty $\sigma_A \times \sqrt{N} \approx \sqrt{2}$;
{\bf (d)}
the approximate value of the uncertainty $\sigma_{\varphi_0} \times \sqrt{N} \approx 1/\sqrt{2}$;
{\bf (e)}
the estimated value (unity) of the the uncertainty normalized to the expected value $\sigma_A \times \sqrt{N} /\sqrt{2-(A \times P)^2}$.
The dashed curve shows the high energy asymptotic expression for $A$
from Ref.\,\cite{Boldyshev:1972va} and eq.\,\ref{eq:sig:HE} and 
the dashed line shows the high-energy asymptotic limit $A = 1/7$ 
{\bf (a)}.
These results were obtained using simulated samples with $P=1$ and $N = 10^5$ events each.
Please note that the same quantity (for example $A$) is measured for various definitions of the azimuthal angle using the same event sample at a given photon energy, so that their statistical fluctuations are correlated.
The error bars in plot {\bf (a)} amount to $\approx \sqrt{2/N} \approx 0.0045$ and are therefore not visible.
 \label{fig:angles}
 }
\end{figure}

The results of measurements of $A$ and of $\varphi_0$ using the above
weights are shown in Figure~\ref{fig:angles}.
The calculation of the uncertainties for $A$ and for $\varphi_0$ are
detailed in \ref{sec:appendix}.

The uncertainty of the measurement of $A \times P$ is
 (eq.\,6 of Ref.\,\cite{Bernard:2013jea}) 
\begin{equation}
 \sigma_{A \times P} \approx \sqrt{\gfrac{2-(A \times P)^2}{N}},
\label{eq:uncertainty:AP}
\end{equation}
where $N$ is the number of events, so the precision of the measurement of $P$ for a cosmic source is 
\begin{equation}
 \sigma_{P} = \gfrac{\sigma_{A \times P}}{A} \approx \gfrac{1}{A}\sqrt{\gfrac{2-(A \times P)^2}{N}},
\label{eq:uncertainty:P}
\end{equation}
that is, $\sigma_{P}$ depends on the definition of the azimuthal angle,
$\varphi$, of the event only through the value of the polarization
asymmetry at photon energy $E$ and measured with that definition of
$\varphi$.
Here we are ``calibrating'' the detector, that is, we simulate an event
sample with a known value of $P$, and we measure $A$ with an
uncertainty
\begin{equation}
 \sigma_{A} = \gfrac{\sigma_{A \times P}}{P} \approx \gfrac{1}{P}\sqrt{\gfrac{2-(A \times P)^2}{N}},
\label{eq:uncertainty:A}
\end{equation}
\begin{itemize}
\item The variation of $A$ with the total available kinetic energy in
 the final state (that is, with $E - 2m c^2$), plot (a),
 shows that the best performance is achieved with the
variable $\phi$ that takes place in the expression of the differential cross
section and that using other expressions only degrades
the precision of the measurement.
\item 
The low and high energy asymptotes for $\varphi \equiv \phi$ are found
to be in good agreement with that predicted by the low energy limit
(eq.\,\ref{eq:sig:LE:final})
and the high energy expression 
(eq.\,\ref{eq:sig:HE}).
\item 
Plot (c) shows that the uncertainty of the measurement of $A$ does not
depend on the definition for $\varphi$ and that it is close to
$\sqrt{2/N}$.
The decrease due to the $A \times P$ dependence of $\sigma_{A \times P}$ 
(eq.\,\ref{eq:uncertainty:AP})
is visible at low energy.
\item
 Plot (e) shows the measured uncertainty of the measurement of
 $A$, normalized to the prediction of eq.\,\ref{eq:uncertainty:AP}
 and confirms that eq.\,\ref{eq:uncertainty:AP} is a good
 representation of the data.
\item Plot (b) confirms that the measurement of the polarization angle
 is unbiased and plot (d) confirms the value of its uncertainty,
\begin{equation}
 \sigma_{\varphi_0} \approx \gfrac{1}{A\sqrt{2 N}}.
\label{eq:uncertainty:varphi0}
\end{equation}

\end{itemize} 

The results presented in this section were obtained from simulations
with the HELAS calculation of the differential cross section. 
(plots not shown).

\clearpage

\section{Conclusion}
\label{sec:conclusion}

We re-examine the polarization asymmetry of $\gamma$ conversion to
$e^+e^-$ pairs with a special focus on the lower part of the energy
range where most of the statistics lies for cosmic sources and for
which gas-detector-based polarimeters presently under development show
very good prospects.
We use an exact event generator of the 5D differential cross section,
either in its analytic Bethe-Heitler form or computed including all
the Feyman diagrams, to obtain event samples at given photon energies
and for either fully polarized or unpolarized photon beams.
We measure the polarization asymmetry of pair conversion as a function
of photon energy from the polarized samples, with a moments method
that use sample averages of trigonometric functions of the event
azimuthal angle.
We demonstrate that defining the azimuthal angle $\varphi$ of the final
state as the bisector angle of the pair, $\phi$, provides the most
precise 1D measurement of the linear polarization fraction of the
incident $\gamma$ radiation.
We find that the polarization asymmetry $A$ keeps on increasing at low
photon energies and reaches a value of $\pi/4$ at threshold.
When the measurement is performed with $\varphi \equiv \phi$, the value
obtained from simulation tends towards that limit at low energies and
is well represented by the high-energy asymptotic expression for 
$E \ge 30\,\mega\electronvolt$.
We confirm the expression of the uncertainty of the measurement of $P$
down to threshold where the value of $A$ is large enough that the
$A$-dependence of $\sigma_{A \times P}$ becomes sizable.
We show that the 5D moments method that makes use of the whole set of
variables that describe the final state also yields an asymmetry that
increases at low energies, and that the gain in precision with respect
to the 1D methods is still sizeable down to threshold.

\section{Acknowledgments}

It is a pleasure to acknowledge the support of the French National
Research Agency (ANR-13-BS05-0002).

\appendix
\section{Measurement of $A$ and of $\varphi_0$: uncertainties}
\label{sec:appendix}

The statistical uncertainty of the measurement of the weight average
$\langle w \rangle$ is obtained from the R.M.S. of the distribution of
$w$: $\sigma(\langle w \rangle) = RMS(w)/\sqrt{N}$.
Propagating the uncertainties using Taylor extensions we obtain:
\begin{eqnarray}
\sigma(A)&=& 2 \gfrac
   {\langle\cos{2\varphi}\rangle \sigma(\cos{2\varphi}) + \langle \sin{2\varphi}\rangle \sigma (\sin{2\varphi})}
   {\sqrt{\langle\cos{2\varphi}\rangle^2+ \langle \sin{2\varphi}\rangle^2}}
\label{eq:stat:sigmaA}
\end{eqnarray}
and
\begin{eqnarray}
  \sigma(\varphi_0)&=&\gfrac{1}{2}\gfrac{\sigma(\sin{2\varphi})\langle \cos{2\varphi}\rangle + \sigma(\cos{2\varphi})\langle \sin{2\varphi}\rangle}{\langle \cos{2\varphi}\rangle^2+\langle \sin{2\varphi}\rangle^2}
.
\label{eq:stat:sigmaPhi0}
\end{eqnarray}

\clearpage

\tableofcontents

\end{document}